\documentstyle[12pt]{article}

\def\rf#1{(\ref{eq:#1})}
\def\lab#1{\label{eq:#1}}
\def\nonu{\nonumber}
\def\br{\begin{eqnarray}}
\def\er{\end{eqnarray}}
\def\be{\begin{equation}}
\def\ee{\end{equation}}
\def\eq{\!\!\!\! &=& \!\!\!\! }

\def\({\left(}
\def\){\right)}

\newcommand{\ct}[1]{\cite{#1}}
\newcommand{\bi}[1]{\bibitem{#1}}

\relax

\def\v{\vert}

\newcommand{\partder}[2]{{{\partial {#1}}\over{\partial {#2}}}}

\newcommand{\sbr}[2]{\left\lbrack\,{#1}\, ,\,{#2}\,\right\rbrack}

\def\d{\delta}

\def\o{\over}

\def\P{\Phi}
\def\pa{\partial}
\def\pr{\prime}

\def\ra{\rightarrow}

\def\tp0{\Theta_{+}^{(0)}}
\def\tm0{\Theta_{-}^{(0)}}

\def\u2{\mid u\mid^2}

\def\vp{\varphi}

\def\ck{{\cal K}}
\def\cl{{\cal L}}

\font \msb=msbm10 scaled \magstep1
\newcommand{\IR}{\mbox{\msb R} }

\newcommand{\IZ}{\mbox{\msb Z} }
\def\one{\hbox{{1}\kern-.25em\hbox{l}}}
\def\0#1{\relax\ifmmode\mathaccent''7017{#1}%
        \else\accent23#1\relax\fi}

\def\PRL#1#2#3{{\sl Phys. Rev. Lett.} {\bf#1} (#2) #3}
\def\NPB#1#2#3{{\sl Nucl. Phys.} {\bf B#1} (#2) #3}

\def\PRD#1#2#3{{\sl Phys. Rev.} {\bf D#1} (#2) #3}

\def\PLB#1#2#3{{\sl Phys. Lett.} {\bf #1B} (#2) #3}

\begin{document}

\begin{titlepage}
\vspace*{-2 cm}
\noindent
April, 1999 \hfill{IFT-P.038/99}\\
hep-th/9905079 \hfill{UICHEP-TH/99-4}

\vskip 3cm

\begin{center}
{\Large\bf  Exact static soliton solutions of ${\bf 3+1}$  dimensional\\ 
integrable theory with nonzero Hopf numbers}
\vglue 1  true cm
H. Aratyn$^1$ , L.A. Ferreira$^2$ and A.H. Zimerman$^2$\\

\vspace{1 cm}
$^1${\footnotesize Department of Physics \\
University of Illinois at Chicago\\
845 W. Taylor St., Chicago, IL 60607-7059}

\vspace{1 cm}

$^2${\footnotesize Instituto de F\'\i sica Te\'orica - IFT/UNESP\\
Rua Pamplona 145\\
01405-900, S\~ao Paulo - SP, Brazil}\\

\medskip
\end{center}

\normalsize
\vskip 0.2cm

\begin{abstract}

In this paper we construct explicitly an infinite number of Hopfions (static, 
soliton solutions with non-zero Hopf topological charges)
within the recently proposed $3+1$-dimensional, integrable and 
relativistically invariant field theory.

Two integers label the family of Hopfions we have found. Their product is equal to 
the Hopf charge which provides a lower bound to the soliton's finite energy.
The Hopfions are constructed explicitly in terms of the toroidal coordinates
and shown to have a form of linked closed vortices.

\end{abstract}
\end{titlepage}

Recent numerical studies of the Faddeev-Skyrme modified $O(3)$ sigma-model
\ct{F-S} support 
existence of static toroidal solutions stabilized by their
non-zero Hopf numbers \ct{numerical}.
The emerging string-like structures are quite intriquing and
may find applications in various physical models of condensed matter physics 
and gauge field theory.

In \ct{AFZ99a} we have introduced the three-dimensional field model which
falls into a class of higher dimensional integrable models from the point of
view of the generalized zero-curvature 
approach \ct{afg}.
The question posed in \ct{AFZ99a} was whether this form of integrability 
is linked to the existence of soliton solutions as it is expected from 
the study of two-dimensional integrable models.
Our analysis of the model in \ct{AFZ99a} has indeed revealed one non-trivial
soliton solution described by a standard Hopf map of unit Hopf index.
To fully establish connection between integrability and soliton solutions would
require finding other topological solitons with arbitrary topological charges.
This is accomplished in this letter. The equations of motion of the model
are solved in toroidal coordinates and the space of solutions 
is found to be represented by a family of
maps $\IR^3 \to \IR^2$ labeled by two integers. 
The integers count number of times the map winds around two independent
angular directions. 

The model under consideration is described by the Lagrangean density
\be
\cl = - \eta_0 \( H_{\mu\nu}^2\)^{3\o 4}
\lab{nicemodel}
\ee
where $\eta_0 = \pm 1$,
determines the choice of the signature of the Minkowski metric, $g_{\mu\nu}
= \eta_0 \,{\rm diag}\( 1, -1,-1,-1\)$
and the field-tensor $H_{\mu\nu}$ is defined in terms of the three component,
unit vector field ${\vec n} \in S^2 $ as
\be 
H_{\mu\nu} \equiv {\vec n} \cdot \( \pa_{\mu} {\vec n} \times \pa_{\nu}{\vec
n}\) \; . 
\lab{hmn}
\ee
The action in \rf{nicemodel} is $O(3)$ and Poincare invariant.
The value $3/4$ of the power of $H_{\mu\nu}^2$ in \rf{nicemodel} is such 
that the theory circumvents the usual
obstacle of Derrick's  scaling argument against existence of stable solitons.
We are interested in the boundary condition ${\vec n}= (0,0,1)$ at spatial
infinity. This condition compactifies effectively the Euclidean space $\IR^3$ 
to the three-sphere $S^3$.
Accordingly, ${\vec n}$ becomes a map: $ S^3 \to S^2$.
Due to $\pi_3 ( S^2) = \IZ$, the field configurations fall into disjoint
classes characterized by the value of the Hopf invariant $Q_H$.

Using  stereographic projection of $S^2$ :
\be
{\vec n} = {1\o {1+\mid u\mid^2}} \, \( u+u^* , -i \( u-u^* \) , \mid u\mid^2
-1 \) 
\lab{stereo}
\ee
one obtains 
\be 
H_{\mu\nu} = {-2 i \o \({1+\u2}\)^2} \, \( \pa_{\mu}u \pa_{\nu}u^* - 
\pa_{\nu}u \pa_{\mu}u^*\)
\lab{hmnu}
\ee
where $u$ is a complex scalar field. 

In terms of the vector quantities  
\be
K_{\mu} =  \frac{i}{2} \({1+\u2}\)^2 \, 
H_{\mu\nu} \pa^{\nu}u = 
\( \pa^{\nu} u^* \pa_{\nu} u \) \, \pa_{\mu} u - 
\(\pa_{\nu} u \)^2 \, \pa_{\mu} u^*
\lab{kmu}
\ee
and
\be
 \ck_{\mu} \equiv { \( K \pa u^* \)^{-{1\o 4}} K_{\mu} \o {1+\u2}}
\ee
the corresponding equations of motion can be rewritten compactly as 
\be
\pa^{\mu}\ck_{\mu} = 0
\lab{eqmot2}
\ee
The quantity \rf{kmu} automatically satisfies the relations 
\be
K^{\mu} \pa_{\mu} u = 0 \qquad \qquad {\rm Im}\( K^{\mu} \pa_{\mu} u^*\) = 0 
\lab{nicerel}  
\ee
which play a crucial role in establishing integrability of the model. Indeed,
using 
relations \rf{nicerel} and the equations of motion \rf{eqmot2} one obtains an 
infinite number of conserved currents given by 
\be
J_{\mu} \equiv  \ck_{\mu}\, {\d G \o \d u} - 
 \ck_{\mu}^*\, {\d G \o \d u^*} 
\lab{infcur2}
\ee
with $G$ being any functional of $u$ and $u^*$ only (no derivatives).
In reference \ct{AFZ99a} we have analyzed
the integrability properties of this theory using the
generalized version of the zero curvature condition \ct{afg}. The equations of
motion \rf{eqmot2} can be represented as 
$\pa^{\mu} B_{\mu}+ \sbr{A^{\mu}}{B_{\mu}}=0$, with $A^{\mu}$ being a flat
$SU(2)$ connection, and $B_{\mu}$ being an operator living in any integer spin
representation of $SU(2)$ (for more details see \ct{AFZ99a,afg}). The
integrability properties emerge due to the fact that $B_{\mu}$ can be put in
any integer spin representation, and that is a direct consequence of 
\rf{nicerel} \ct{afg,fl}.
The question we address here is whether the above notion of integrability
implies existence of the infinite many soliton solutions.

We shall look for time-independent solutions to equations of motion
\rf{eqmot2} of the type\footnote{We shall take $m$ and $n$ integers in order
for $u$ to be single valued.}:
\be
u \( \eta, \xi, \vp \) \equiv f (\eta) e^{ i  (m \xi+ n \vp)}
\lab{genu}
\ee
where we used toroidal coordinates on $\IR^3$:
\br
x &=& a q^{-1} \sinh \eta \cos \vp \;\;, \;\;
y = a q^{-1} \sinh \eta \sin \vp   \nonu \\
z &=& a q^{-1} \sin \xi \quad ; \quad a > 0 \;\; ; \; q = \cosh \eta - \cos \xi
\lab{tordefs}
\er
The angles $\xi , \vp$ both vary from $0$ to $2 \pi$ and $\eta$ varies from $0$
to $\infty$. The surfaces of constant $\eta$ are toroids that circle
the $z$-axis, $\xi$=constant are spheres and $\vp$=constant are half-planes.
The corresponding gradient of $u$ becomes in toroidal coordinates:
\be
{\vec \nabla} u = (q/a) e^{i  (m \xi+ n \vp)}  \( f^{\pr} (\eta) {\hat e}_{\eta} 
+ i m f (\eta) {\hat e}_{\xi} +i n f (\eta) { {\hat e}_{\vp} \o \sinh \eta} \, \)
\lab{gradbu}
\ee
where we introduced the unit vectors spanning the orthogonal toroidal 
coordinate system having properties 
${\hat e}_{i} \cdot {\hat e}_{j} =\d_{ij}$, $i,j \equiv \eta ,\xi , \vp$.  

In terms of the vector-field ${\vec \ck} \equiv 
\( {\vec K} \cdot {\vec \nabla} u^*\)^{-1/4} {\vec K}
 / (\v u \v^2 +1)$
the equations of motion \rf{eqmot2} become, in the static case,
${\vec \nabla} \cdot {\vec \ck} =0$, i.e. ${\vec \ck}$ is solenoidal.
Using \rf{genu} and \rf{gradbu} we obtain for the components 
of ${\vec \ck}$:
\be
\ck_{\eta} = \ck_0  \( m^2 + { n^2 \o (\sinh \eta)^2} \) \, f \; ;\qquad 
\ck_{\xi}= i  \, m \ck_0  \, f^{\pr} \; ; \qquad 
\ck_{\vp} = i \, n \ck_0  \, {f^{\pr}\o \sinh \eta}
\lab{vpbu}
\ee
with 
\be
\ck_0  \equiv \sqrt{2} (q/a)^2  e^{ i  (m \xi+ n \vp)} 
\( m^2 + { n^2 \o (\sinh \eta)^2} \)^{-1/4} 
{ f^{1/2} f^{\pr \, 1/2} \o (f^2 +1)} 
\lab{vpbu2}
\ee
where the components of the vector field are defined according to
$ {\vec V} =  {\hat e}_\eta V_\eta + {\hat e}_{\xi}
V_{\xi}+ {\hat e}_{\vp}V_{\vp}$.

Plugging the components of ${\vec \ck}$ into the expression for divergence
of the vector field in the toroidal coordinates
\be
{\vec \nabla} \cdot {\vec V} = {q \o a} \, \( \partder{V_\eta}{\eta}
+ \partder{V_{\xi}}{\xi} + {1 \o \sinh \eta}\, \partder{V_{\vp}}{\vp}
- 2 V_{\xi}\, { \sin \xi \o q} + V_\eta \, ( {\cosh \eta \o \sinh \eta } -
{ 2 \sinh \eta \o q}) \)
\lab{divtoro}
\ee
and using 
\be
\partder{q^2}{\eta}= q^2 {2 \sinh \eta \o q} \quad ;\quad 
\partder{q^2}{\xi}= q^2 {2 \sin \xi \o q} 
\lab{qsqunu}
\ee
we arrive at:
\be
\partder{}{\eta} \ln { f f^{\pr} \o (f^2 +1)^2}= - { 2 m^2 (\sinh \eta)^2 -n^2
\o m^2 (\sinh \eta)^2 +n^2 } { \cosh \eta \o \sinh \eta}
\lab{bueqsofmo}
\ee
We take $m^2 > n^2$  in eq.\rf{bueqsofmo}.
The integration yields:
\be
{1 \o f^2 +1 } = {2 k_1 \o \v m \v (m^2 -n^2)} { \cosh \eta  
  \o \( {n^2 - m^2 \o m^2} +
\cosh^2 \eta \)^{1/2} } +k_2
\lab{fspone}
\ee
with $k_2$ being the integration constant of the last integration.
Imposing boundary conditions:
\br
{\vec n} &\ra& \( 0 , 0, 1 \) \qquad {\rm or} \qquad \mid u \mid \ra \infty 
\qquad {\rm or} \qquad f \ra \infty \qquad {\rm as} \qquad \eta \ra 0 \\
{\vec n} &\ra& \( 0 , 0, -1 \) \qquad {\rm or} \qquad u \ra 0  
\qquad {\rm or} \qquad f \ra 0 \qquad {\rm as} \qquad \eta \ra \infty 
\lab{boundary}
\er
one gets
\be
f^2 = {   \cosh \eta  - \sqrt{n^2/m^2 +\sinh^2 \eta}  \o
 \sqrt{1 + m^2/n^2\,\sinh^2 \eta  }  -   \cosh \eta  }
\lab{fsquare}
\ee
One observes that $f$ depends only on the ratio $m^2/n^2$,  and that $f^2 \geq
0$ 
for any value of $m^2/n^2$.  
Note, that taking the limit $m \to n$ in \rf{fsquare} and using L'Hospital
rule yields:
\be
\lim_{m \to n} f^2 = {1 \o \sinh^2 \eta}
\lab{mnfsq}
\ee
which gives for $m=n$ a solution:
\be
u = {e^{im (\xi + \vp)} \o \sinh \eta}
\lab{bumen}
\ee
In the special case of $m=n=1$ eq.\rf{bumen} reproduces the standard soliton
solution  
\ct{nicole,AFZ99a}. Expression \rf{bumen} can be written as a composite of the
Hopf  
map together with the stereographic map $:\IR^3 \to S^3$ of degree $1$.

A formula:
\be
E \equiv \int d^3x \, \Theta_{00} = 
 8^{3\o 4}\, \int d^3x \,  { \( K_{i} \pa^{i} u^* \)^{3\o 4} \o  
\( 1+\mid u \mid^2\)^3}
\lab{energy}
\ee
describes energy of the static configuration \ct{AFZ99a}.
Inserting our solution into \rf{energy} we obtain an expression
for the energy $E=E_{m,n}$ of the soliton configuration 
\be
E_{m,n} = (2 \pi)^2 8 \cdot 2^{3/4} \int_0^{\infty} {d\eta \sinh \eta \o 
  (1+ f^2)^3}
\( m^2 + { n^2 \o (\sinh \eta)^2} \)^{3/4}  f^{3/2} f^{\pr \, 3/2}
\lab{emn}
\ee
which after the $\eta$-integration yields:
\be
E_{m,n} =  (2 \pi)^2 4 \cdot 2^{1/4} \, \sqrt{\v n \v \v m \v  (\v n \v +\v m
\v)} 
\lab{emnb}
\ee

We now turn to calculation of the Hopf numbers.
Define functions $\P_i, i=1, {\ldots}, 4$ as follows:
\br
\P_{\left\{ \begin{array}{c} 1\\
2 \end{array} \right\} } \eq \( {   \cosh \eta - \( n^2/m^2+ \sinh^2 \eta
\)^{1/2} \ \o 
  (\v m/n \v -1) \( n^2/m^2+ \sinh^2 \eta \)^{1/2} } \)^{1/2} 
{\left\{\begin{array}{c} \cos m \xi \\ \sin m \xi \end{array}\right\}}
\lab{phione}\\
\P_{\left\{\begin{array}{c} 3\\
4 \end{array}\right\}} \eq \( { \( 1 + (m^2/n^2)\sinh^2 \eta \)^{1/2} -
 \cosh \eta \o 
  (\v m/n \v  -1) \( n^2/m^2+ \sinh^2 \eta \)^{1/2} } \)^{1/2} \,
{\left\{\begin{array}{c} 
  \cos n \vp \\ - \sin n \vp  \end{array}\right\}}
\er
They provide parametrization of $S^3$ and satisfy :
\be
u = { Z_1 \o Z_2} = { \P_1 + i \P_2 \o \P_3 + i \P_4} \qquad \qquad 
\mid Z_1\mid^2 + \mid Z_2\mid^2 =1  
\lab{uzz}
\ee
or equivalently, for ${\vec n}$ defined from $u$ via relation \rf{stereo} :
\be
n_i = Z^{\dag} \sigma_i Z 
\quad ; \quad Z^{\dag} = \( Z_1^*, Z_2^* \) \; ; \; 
Z =  {Z_1 \choose Z_2}
\lab{nsigma}
\ee
Let, furthermore 
\be
A_i = {i \o 2} \( Z^{\dag} \pa_i Z - \pa_i Z^{\dag}\,  Z \)
\lab{vec-pot}
\ee
be a vector potential for the two-form $H_{ij}= \pa_i A_j -\pa_j A_i$.
In terms of the toroidal coordinates the components of ${\vec A}$ are
\be
A_{\eta} = q \pa_{\eta} \( \P_{3} \P_4 \) \; ; \; 
A_{\xi} = - m q \( \P_1^2 + \P_2^2\) \; ; \; 
A_{\vp} = - {n q \o \sinh \eta} \( \P_3^2 + \P_4^2\)
\lab{ap}
\ee
Next, we calculate the vector function
${\vec B} = {\vec \nabla} \times {\vec A}$
(such that $B_i = \epsilon_{ijk} H_{jk} /2$),
using expression for the curl operator in toroidal coordinates
and plugging it into the Hopf index defined as :
\be
Q_H = {1 \o 4 \pi^2} \int d^3 x\, {\vec A} \cdot {\vec B}
\lab{qhab}
\ee
for which we now find
\be
Q_H =  { n m  \o 2 }  \, \( \( \P_1^2 + \P_2^2 \)^2 \Bigg\vert_{0}^{\infty} 
-  \( \P_3^2 + \P_4^2 \)^2 \Bigg\vert_{0}^{\infty} \)
=  - nm
\lab{hopfinho}
\ee
It is always possible to choose the Hopf index such that it is positive $Q_H\geq 0$.
This amounts to the right choice of orientation, which determines
the sign of $Q_H$.

Note, that the following inequality:
\be
\v m \v + \v n \v  \geq 2 \sqrt{\v n \v \v m\v }
\lab{ineq}
\ee
always holds.
Hence
\be 
\sqrt{\v n \v \v m \v  (\v n \v +\v m \v)}
 \geq \sqrt{2} (\v n \v \v m\v )^{3/4}
\; \to \; E_{m,n} \geq   (2 \pi)^2 4 \cdot 2^{3/4} \, \v Q_H \v^{3/4}
\lab{vakulenko}
\ee
which agrees with the lower bound result by \ct{VK79} for the Faddeev-Skyrme model.

Notice that the solution obtained above implies that the vector field 
${\vec n}$ (see \rf{stereo}) is given by  
\be
n_1 = \frac{2f}{f^2+1} \, \cos \( m \xi + n \vp\) \; ; \qquad 
n_2 = \frac{2f}{f^2+1} \, \sin \( m \xi + n \vp\) \; ; \qquad 
 n_3 = \frac{f^2 -1}{f^2+1}
\ee
Since $f$ is a function of $\eta$ only, one observes that the surfaces of
constant $n_3$ are tori. In addition, on those surfaces, the lines of constant
$n_1$ and $n_2$ wind around the tori with frequencies in the $\xi$ and
$\vp$-directions given by $m$ and $n$, respectively. One can check that $n_3$
falls monotonically from $n_3=1$ at $\eta =0$ to $n_3=-1$ at $\eta =
\infty$. In addition, the bigger the ratio $m/n$ is, the faster if performs
that flip. Therefore the size of our soliton decreases with the increase of
$m/n$. 

The Hopf index of our solution can alternatively be calculated in the
following way. The solution provides a mapping of the spatial $\IR^3$ into
$S^3$ defined by $\sum_{i=1}^4 \Phi_i^2 =1$. Then, \rf{stereo} and \rf{uzz}
provide the Hopf map $S^3 \ra S^2$. The Hopf index is given by the linking
number of the pre-images, under the Hopf map, of any two points of $S^2$. 
Consider the point ${\vec n} = \( 0,0,-1\)$ which corresponds in $\IR^3$ to
$\eta \ra \infty$. For this value of $\eta$, $\P_1, \P_2$ go to zero and
$\P_3 = \cos n \vp$ and $\P_4 = - \sin n \vp$. On $S^3$ we find the circle
$\P_3^2+\P_4^2 = \cos^2 n \vp + \sin^2 n \vp = 1$ of radius $1$
and as $\vp$ varies between $0$ and $2 \pi$ the preimage wraps $\mid n\mid$
times around the $\vp$ direction.
Similarly the preimage of ${\vec n} = \( 0,0,1\)$ corresponds to $\eta
=0 $. For this value of $\eta$, $\P_3, \P_4$ go to zero and
$\P_1 = \cos m \xi$ and $\P_2 =  \sin m \xi$. On $S^3$ we find the circle
$\P_1^2+\P_2^2 = \cos^2 m \xi + \sin^2 m \xi = 1$ of radius $1$
and as $\xi$ varies between $0$ and 
$2 \pi$ the preimage wraps $\mid m\mid$ times around the $\xi$ direction.
Since these two circles intersect the linking number is $\mid nm\mid $.
That is   indeed the Hopf number calculated above.

Finally, let us mention that other models circumventing  Derrick's theorem
have been  
proposed previously in the literature \ct{deser,nicole}. A common feature is
that, like 
the model considered in the present paper, the Lagragians are non-polynomial
functions of the fields and their derivatives. The 
corresponding solutions have been constructed and in the special case of
\ct{nicole} a soliton  with the unit Hopf charge was obtained.
To our knowledge, an infinite number of soliton solutions
to the field theoretical equations of motion with the Hopf charges bigger than 
unity were not obtained previously in the literature in an exact form. 
 
\vspace{1 cm}

\noindent {\bf Acknowledgements}\\
LAF and AHZ are partially supported by CNPq (Brazil).

\end{document}